\documentclass[aps,prl,reprint,superscriptaddress]{revtex4-2}

\usepackage[T1]{fontenc}
\usepackage{amsmath,amssymb}
\usepackage{graphicx}
\usepackage{bm}

\begin{document}

\title{Emergent Yielding from Structural Load Transfer in Disordered Soft Solids}
\author{Lalit Kumar}
\email{lalit.kumar@iitb.ac.in}
\affiliation{Department of Energy Science and Engineering, Indian Institute of Technology Bombay, Mumbai 400076, India}

\date{\today}

\begin{abstract}
Yielding in disordered soft solids originates from the progressive redistribution of load-bearing capacity from recoverable elastic networks to frictional interactions through deformation-induced structural evolution. We present a unified, non-singular constitutive framework demonstrating that this mechanism naturally generates a finite yield stress and a smooth solid-to-fluid transition without prescribed yield criteria, constitutive switching, or divergent viscosities. The framework captures diverse transient and steady rheological phenomena, including Herschel–Bulkley behavior, stress overshoots, hysteresis, viscosity bifurcation, plug-flow formation, and thixotropic steady-state shear banding.

\end{abstract}
\maketitle

Disordered soft solids, such as colloidal gels, foams, and emulsions, undergo a driven nonequilibrium transition from solid-like elasticity to liquid-like flow. The physical origin of yielding in these materials remains a longstanding unresolved question. Although macroscopic signatures of this transition, including yielding, shear thinning, and transient localization, are well documented~\cite{Bonn2017,Weeks2007}, a predictive physical mechanism linking deformation-induced structural evolution to the emergence of a finite yield stress remains largely elusive.  Classical macroscopic formulations~\cite{Bingham,HB} rely on piecewise constitutive rules that introduce singular apparent viscosities in the limit of 
vanishing shear rate ($\dot{\gamma}\rightarrow0$). Conversely,  
nonlinear viscoelastic frameworks~\cite{OldroydB,FENEP,PTT,Giesekus1982} 
regularize stress growth but fail to predict a finite residual stress in the 
hydrodynamic limit. Modern statistical and structural approaches---including Soft Glassy Rheology~\cite{Sollich1997}, fluidity models~\cite{Fluidity,HebraudLequeux}, aging and fluidization models~\cite{Derec2001,Varnik2003}, thixotropic constitutive models~\cite{Coussot2002, MewisWagner, deSouzaMendes2012}, and more recent structure-dependent constitutive formulations~\cite{KumarL}---successfully capture startup stress overshoots, viscosity--bifurcation, and shear localization associated with complex yielding dynamics~\cite{Fielding2014,MoorcroftFielding2013,Olmsted2008,Ovarlez2009,Moller2009}.  However, these descriptions generally rely on structural state variables, kinetic evolution equations, or phenomenological yielding mechanisms, and do not establish thermodynamically consistent physical mechanism by which a finite yield stress emerges. Here, we address this question by developing a thermodynamically consistent framework in which deformation-induced structural evolution redistributes load-bearing capacity to generate a finite yield stress. 

\noindent We present a unified, non-singular constitutive framework in which a finite yield stress emerges from the redistribution of load-bearing capacity. Rather than prescribing a macroscopic yield criterion or introducing divergent viscosities, the proposed mechanism naturally produces a smooth solid-to-fluid transition and captures Bingham and Herschel–Bulkley-type rheology, startup stress overshoots, plug-flow behavior, hysteresis, and viscosity bifurcation. A simple extension incorporating structural rebuilding further predicts non-monotonic constitutive behavior and steady-state shear banding in thixotropic materials~\cite{MoorcroftFielding2013,Fielding2014}. 

\begin{figure}[t]
\centering
\includegraphics[width=0.95\columnwidth]{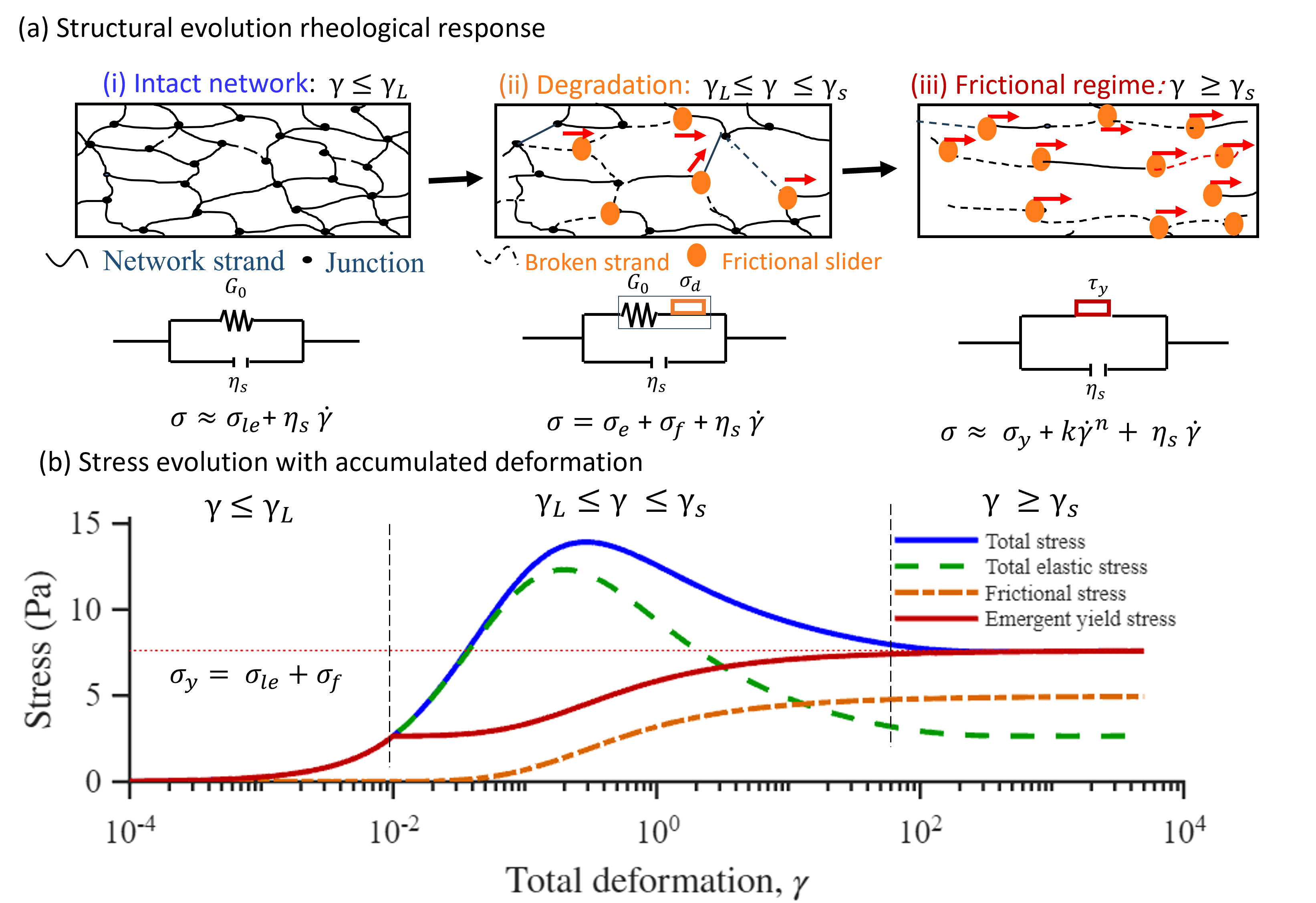}
\caption{\textbf{Physical mechanism of load transfer and emergent yielding.} 
(a) Schematic illustration of the proposed yielding mechanism. Small deformations are stored reversibly within the intact load-bearing network. Beyond the onset of nonlinearity, structural rearrangements progressively reduce nonlinear elastic stress while promoting frictional sliding. At large deformation, frictional interactions saturate, producing a finite emergent yield stress. (b) Model prediction of evolution of the stress contributions with accumulated deformation. The total stress (blue) results from the competition between elastic stress (green) and frictional stress (orange). The emergent yield stress (red) equals the sum of the persistent linear elastic stress and the saturated frictional stress, $\sigma_y=\sigma_{le}+\sigma_f$.}
\label{fig:fig1}
\end{figure}

\textit{Proposed Mechanism for Emergent Yielding}: Figure \ref{fig:fig1} schematically illustrates the mechanism for emergent yielding. As shown in Fig.~\ref{fig:fig1}(a), elastic stresses initially dominate the response and store deformation energy in the load-bearing structure. Small initial linear elastic deformations are stored reversibly and leave the microstructure largely unchanged. At larger deformation, nonlinear elastic loading is accompanied by structural rearrangements that degrade the load-bearing network. Simultaneously, the degraded microstructure promotes frictional interactions, leading to a contribution of frictional stress consistent with observations of plastic deformation and frictional dissipation in soft glassy materials and concentrated suspensions \cite{FalkLanger1998, Mari2014}. With increasing deformation, the frictional stress grows and ultimately approaches a finite saturation value. This saturation may be associated with the exhaustion of deformation-driven structural degradation or the establishment of a dynamic structural equilibrium \cite{Sollich1997, Coussot2002}. The resulting saturated frictional stress, together with the linear elastic stress, gives rise to a finite emergent yield stress. The constitutive equations below provide a quantitative description of this mechanism:

\begin{equation}
\sigma
=
\sigma_e(\gamma_e)
+
\sigma_d(\Lambda_e,\Gamma,\dot{\gamma})
+
\eta_s\dot{\gamma},
\label{eq:eq1}
\end{equation}

\noindent where $\sigma_e$, $\sigma_d$, and $\eta_s$ denote the elastic, frictional, and viscous stress contributions, respectively. Here, $\dot{\gamma}$ is the imposed shear rate and $\Lambda_e$ is the effective interaction strength. The total deformation is denoted by $\gamma$, while $\Gamma = \langle \gamma - \Gamma_L \rangle$ represents the accumulated nonlinear deformation associated with structural rearrangements beyond the characteristic deformation $\Gamma_L$, which marks the onset of nonlinear response. The total deformation is therefore partitioned into a recoverable linear elastic component and a nonlinear deformation component,

\begin{equation}
\gamma = \underbrace{(\gamma - \Gamma)}_{\text{linear elastic deformation}} + \underbrace{\Gamma}_{\text{ total nonlinear deformation}},
\label{eq:eq2}
\end{equation}

\noindent where the linear elastic component stores deformation reversibly with negligible structural degradation, and the nonlinear deformation comprises both nonlinear elastic and unrecoverable contributions. The elastic stress is given by $\sigma_e = G_0 \gamma_e $, where
\begin{equation}
\gamma_e
=\gamma_{le}+\gamma_{nle}=
(\gamma-\Gamma)
+
\frac{\Gamma}{(2\Gamma/\Gamma_c+1)^{3/2}} .
\label{eq:eq3}
\end{equation}

\noindent The recoverable nonlinear deformation decreases continuously with accumulated nonlinear deformation owing to progressive structural rearrangements, consistent with microscopic rearrangements observed at the onset of flow~\cite{MoorcroftFielding2013,Fielding2014}. The presented elastic stress form is derived from a deformation-dependent free energy. The nonlinear elastic weakening is governed directly by deformation history rather than an independent structural variable~\cite{Bautista1999,deSouzaMendes2012,deSouzaMendes2013}. Here, $\phi(\Gamma)=
\left(
1+2\Gamma/\Gamma_c
\right)^{-1/2}$ is introduced as the structural survival factor,
while $\phi^3(\Gamma)=
(
1+2\Gamma/\Gamma_c)^{-3/2}$ represents the surviving nonlinear load-bearing
fraction. The structural survival factor $\phi$ provides a direct measure of the fraction of nonlinear load-bearing capacity that remains active during deformation. The characteristic deformation $\Gamma_c$ controls the
progressive loss of nonlinear load-bearing capacity through
deformation-induced structural rearrangements. The frictional stress is
\begin{equation}
\sigma_d =
3\,G_0\,\Lambda_e(\dot{\gamma})\,\Gamma
\left[
\phi^2(\Gamma)
-
\phi^3(\Gamma)
\right].
\label{eq:eq5}
\end{equation}
This term can be interpreted as arising from a distribution of interaction strengths within the load-bearing structure, where increasing deformation depletes the population of load-bearing elements and promotes frictional sliding between the degraded and stretched structural elements. The chosen form ensures a smooth quadratic growth of frictional stress at small deformation and a finite saturation value at large deformation. After combining all terms, the constitutive model becomes
\begin{align}
\sigma(\gamma_e,\Gamma,\dot{\gamma})
=
G_0\gamma_e
+3\,G_0\,\Lambda_e(\dot{\gamma})\,\Gamma
\left[\phi^2(\Gamma)-\phi^3(\Gamma)
\right]
+
\eta_s \dot{\gamma},
\label{eq:eq6}
\end{align}

\noindent  The model given in Eq.~\ref{eq:eq6} reproduces the sequence of mechanisms proposed schematically in Fig.~\ref{fig:fig1}(a). With increasing deformation, the capacity of the material to sustain recoverable elastic stresses decreases, while the frictional contribution continues to increase toward its saturation value, as illustrated in Fig.~\ref{fig:fig1}(b). Unless otherwise stated, all results are obtained using
$G_0=265~\mathrm{Pa}$,
$\Lambda_{e0}=0.05$,
$K=0.05$,
$n=0.5$,
$\Gamma_L=0.05\Gamma_C$,
$\eta_s=0.005~\mathrm{Pa\,s}$,
and $\Gamma_L+\Gamma_C=0.05$. The interplay between the decaying nonlinear elastic contribution and the saturating frictional contribution together with linear elastic stress therefore produces a finite asymptotic stress identified as the emergent yield stress. Figure~\ref{fig:fig1}(b) further demonstrates that all stress contributions evolve smoothly with $\Gamma$ and remain finite throughout the deformation history. The same mechanism also gives rise to the transient stress overshoots discussed later, with stress initially increasing due to elastic loading before decreasing as irreversible deformation erodes the material's ability to store elastic stress. Unlike classical yield-stress models that invoke an imposed yield criterion, viscosity divergence, or constitutive singularity, the proposed model remains fully non-singular and unified while naturally approaching a finite asymptotic stress. Yielding therefore emerges as a consequence of structural load transfer rather than being prescribed a priori. This result highlights that a unified constitutive framework can generate a finite yield stress while remaining continuously differentiable.

\textit{Thermodynamic consistency}: The total deformation is partitioned into a recoverable elastic deformation, $\gamma_e$, and an irrecoverable structural deformation, $\gamma_i$, such that

\begin{equation}
\gamma=\gamma_e+\gamma_i.
\label{eq:decomp}
\end{equation}

\noindent The total deformation is decomposed into recoverable elastic and irrecoverable structural components. Unlike classical elastoplasticity, structural evolution continuously redistributes deformation between these components while preserving the imposed total deformation.  Consequently, yielding is interpreted as a continuous redistribution of load-bearing capacity rather than an abrupt transition between elastic and plastic states. The Helmholtz free-energy density is assumed to depend only on the recoverable elastic deformation,$\psi(\gamma_e)
=
G_0\gamma_e^2/2,$ where $G_0$ denotes the elastic modulus. Since the irrecoverable structural deformation does not contribute to elastic energy storage, $\partial\psi/\partial\gamma_i=0.$ Applying the Coleman--Noll procedure gives the elastic stress as

\begin{equation}
\sigma_e
=
\frac{\partial\psi}{\partial\gamma_e}
=
G_0\gamma_e.
\label{eq:sigmae}
\end{equation}

\noindent The recoverable elastic deformation is related to the accumulated nonlinear deformation through
which immediately yields the irrecoverable structural deformation,

\begin{equation}
\gamma_i
=
\gamma-\gamma_e
=
\Gamma
-
\Gamma\phi^3(\Gamma) .
\label{eq:gammai}
\end{equation}

\noindent Equation~(\ref{eq:gammai}) reveals that initially, nearly all deformation is recoverable, whereas increasing nonlinear deformation increases the irrecoverable structural deformation and reduces the recoverable load-bearing capacity.
For isothermal deformation, the Clausius--Duhem inequality requires $\mathcal{D}
=
\sigma\dot{\gamma}
-
\dot{\psi}
\ge0.
$ Since the free energy depends only on the recoverable elastic deformation,

\begin{equation}
\dot{\psi}
=
\frac{\partial\psi}{\partial\gamma_e}\dot{\gamma}_e
=
\sigma_e\dot{\gamma}_e,
\end{equation}

\noindent and using the kinematic decomposition $\dot{\gamma}
=
\dot{\gamma}_e+\dot{\gamma}_i,$ the dissipation reduces to

\begin{equation}
\boxed{
\mathcal{D}
=
\sigma_e\dot{\gamma}_i
+
\sigma_f\dot{\gamma}
+
\mu_s\dot{\gamma}^{\,2}
\ge0.
}
\label{eq:dissipation}
\end{equation}

\noindent  The first term represents the irreversible conversion of recoverable elastic deformation into irrecoverable structural deformation, the second term corresponds to frictional dissipation associated with structural rearrangements, and the third term represents viscous dissipation. The present framework therefore satisfies the second law of thermodynamics while providing a unified physical description in which yielding emerges through the continuous transfer of load-bearing capacity from recoverable elastic deformation to irrecoverable structural deformation and ultimately to frictional resistance. (for more details see Supplemental Material)

\begin{figure}[t]
\centering
\includegraphics[width=0.8\columnwidth]{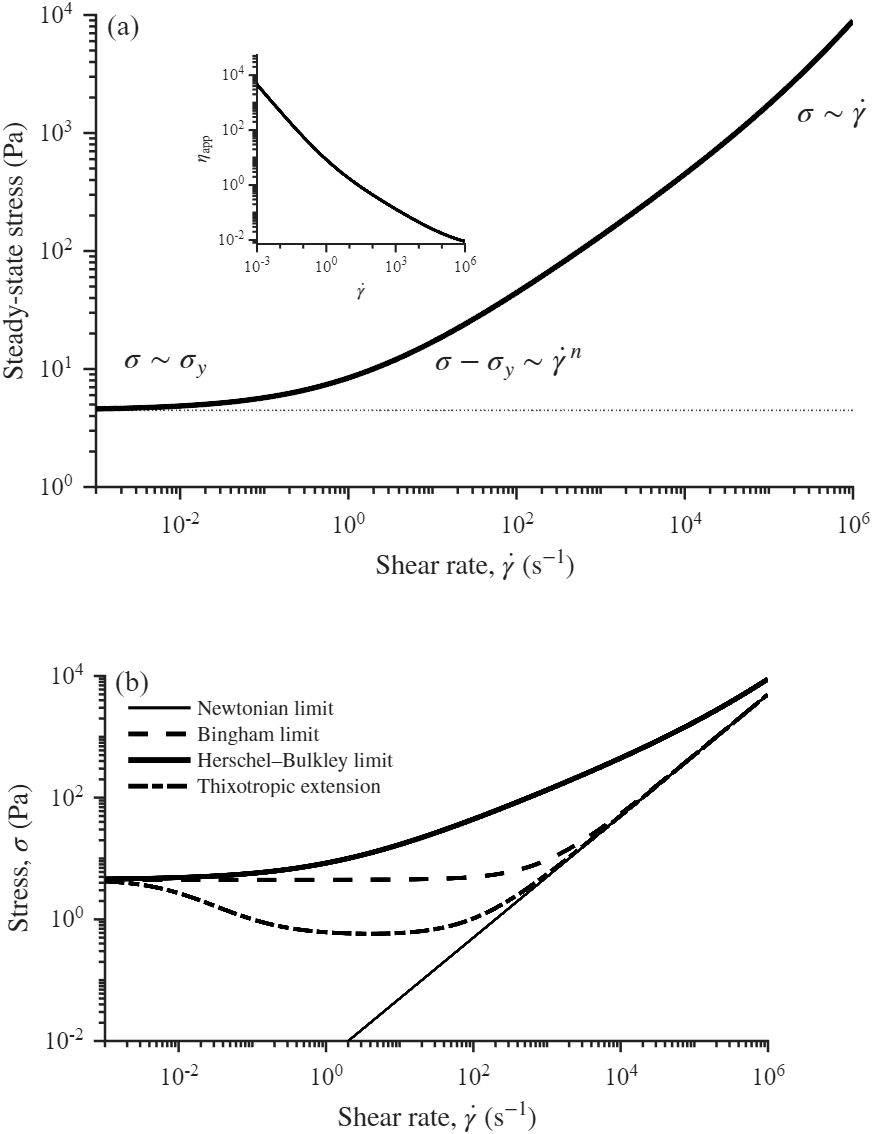}
\caption{\textbf{Steady-state rheology predicted by the load-transfer framework}
(a) Steady-state constitutive response of the rate-dependent load-saturation model. At low shear rates, the stress approaches a finite yield-stress plateau ($\sigma \approx \sigma_y$). At intermediate shear rates, the model exhibits Herschel--Bulkley behavior ($\sigma-\sigma_y \sim \dot{\gamma}^{n}$), while at high shear rates the Newtonian limit ($\sigma \sim \eta_s\dot{\gamma}$) is recovered. Inset: apparent viscosity ($\eta_{\rm app}=\sigma/\dot{\gamma}$).
(b) Hierarchy of constitutive responses generated. Constant frictional interactions produce a Bingham-like response, rate-dependent frictional interactions produce Herschel--Bulkley shear thinning, and thixotropic structural evolution ($b=100{\rm s}$) generate a non-monotonic constitutive response satisfying the classical criterion for steady-state shear banding. Here, $\Gamma_L=0.01\Gamma_c$ is considered.
}
\label{fig:fig2}
\end{figure}

\textit{Emergent Yield Stress and Steady-State Rheology}: In contrast to conventional viscoelastic constitutive models such as Johnson--Segalman, Giesekus, FENE-P, and Rolie--Poly~\cite{Johnson1977,Giesekus1982,Bird1980,Likhtman2003}, which may exhibit stress overshoots and apparent yield-like behavior but possess no finite stress intercept as $\dot{\gamma}\rightarrow0$, the present model predicts a finite yield stress. The finite intercept arises from saturation of the frictional contribution together with stored linear elastic stress [Fig.~\ref{fig:fig2}(a)], the large-deformation limit $(\Gamma\to \infty)$ yields
\begin{equation}
\sigma_e \to G_0\Gamma_L, \quad
\sigma_y \to \sigma_d =G_0\Gamma_L+3G_0 \Lambda_{e0}\Gamma_c/2.
\label{eq:eq9}
\end{equation}
\noindent Equation (11) reveals the central mechanism of the model: nonlinear elastic stress storage vanishes under sustained deformation, whereas the frictional stress approaches a finite saturation value. The yield stress is therefore an emergent consequence of structural load-transfer rather than a prescribed constitutive property. Furthermore, unlike elastoviscoplastic models that impose yielding through a prescribed yield criterion~\cite{Saramito2007}, the present model allows the solid-to-fluid transition to emerge continuously without constitutive switching. 

\noindent The rate-dependent interaction strength is assumed to depend on $K\dot{\gamma}^n$, which characterizes sliding induced rate-dependent energy dissipation. The power-law form, $\Lambda_e=\Lambda_{e0}+K\dot{\gamma}^n,$ is the simplest choice consistent with the experimentally observed Herschel--Bulkley scaling. Here, $\Lambda_{e0}$ denotes the interaction strength in the limit of vanishing shear rate.  In the steady-state limit $(\Gamma\to \infty)$ the model reduces to
\begin{equation}
\sigma = \sigma_y + C \dot{\gamma}^n + \eta_s \dot{\gamma},
\label{eq:eq11}
\end{equation}
\noindent where, $C = 3 G_0  \Gamma_cK/2$. Three distinct flow regimes arise from Eq.~(12): a yield-stress-dominated regime at low shear rates, a Herschel--Bulkley regime at intermediate shear rates, and a Newtonian regime at sufficiently high shear rates. The intermediate and high-shear-rate asymptotic behaviors are highlighted in Fig.~\ref{fig:fig2}(a) using $\sigma-\sigma_y\sim\dot{\gamma}^{\,n}$ and $\sigma\sim\dot{\gamma}$, respectively. Thus, the apparent steady-state viscosity diverges as $\dot{\gamma}\rightarrow 0$, decreases through an extended shear-thinning regime, and asymptotically approaches the solvent viscosity at high shear rates [inset Fig.~\ref{fig:fig2}(a)].

 A significant consequence of the structural load-transfer framework is that steady-state non-monotonic constitutive behavior emerges naturally when the interaction strength is coupled to a thixotropic structural parameter, represented here by $\Lambda_e/(1+b\dot{\gamma})$.  Such non-monotonic flow curves are often obtained only through considerably more elaborate thixotropic, fluidity, or thixo-elasto-viscoplastic formulations and are commonly associated with steady-state shear banding and flow heterogeneity~\cite{Coussot2002,KumarL,MoorcroftFielding2013, Fielding2014,Dimitriou2014}. Time-dependent rebuilding restores the load-bearing elastic network and captures aging and recovery effects (see Supplemental Material). 
Differentiating  the steady-state model ($\Gamma\to \infty$) without thixotropy, Eq.~(5), gives
\begin{equation}
\frac{d\sigma}{d\dot{\gamma}}
=
\frac{3G_0  \Gamma_cK n}{2} \dot{\gamma}^{\,n-1}
+
\eta_s.
\end{equation}
as $K>0$, $0<n<1$, and $\eta_s>0$, the above equation yields $d\sigma/d\dot{\gamma} > 0$. Thus, the frictional-saturation model does not produce steady-state instability or shear banding. This result demonstrates that frictional saturation alone generates a finite emergent yield stress while maintaining a mechanically stable constitutive response. Steady-state shear localization therefore requires an additional mechanism associated with structural rebuilding, introduced in the thixotropic extension. Figure~\ref{fig:fig2}(b) illustrates how increasingly complex rheological behavior emerges naturally within the load-transfer framework. Constant frictional interactions produce a Bingham behavior, while rate-dependent frictional activation generates Herschel--Bulkley-like shear thinning. Incorporating structural rebuilding introduces a competition between the rebuilding and flow timescales. At low shear rates, rebuilding maintains a larger steady-state interaction strength, increasing the frictional stress contribution. When this increase overtakes viscous stress, a non-monotonic constitutive response emerges ( $d\sigma/d\dot{\gamma} < 0$), leading to constitutive instability and steady-state shear banding  (see Supplemental Material). The resulting framework therefore provides a unified description of both simple and thixotropic yield-stress fluids.

\textit{Transient Flow}: The minimal constitutive framework quantitatively reproduces startup shear experiments of bentonite over a wide range of shear rates using a single set of parameters (see Supplemental Material, Fig. S1). The deviation of the linearity parameter, $\Gamma_L$, shows a week dependent on shear rates, and it is varied based on experimental data.. The following analysis employs the minimum constitutive model to reveal the fundamental mechanisms underlying transient stress evolution and emergent yielding.

\begin{figure}[t]
\centering
\includegraphics[width=1.0\columnwidth]{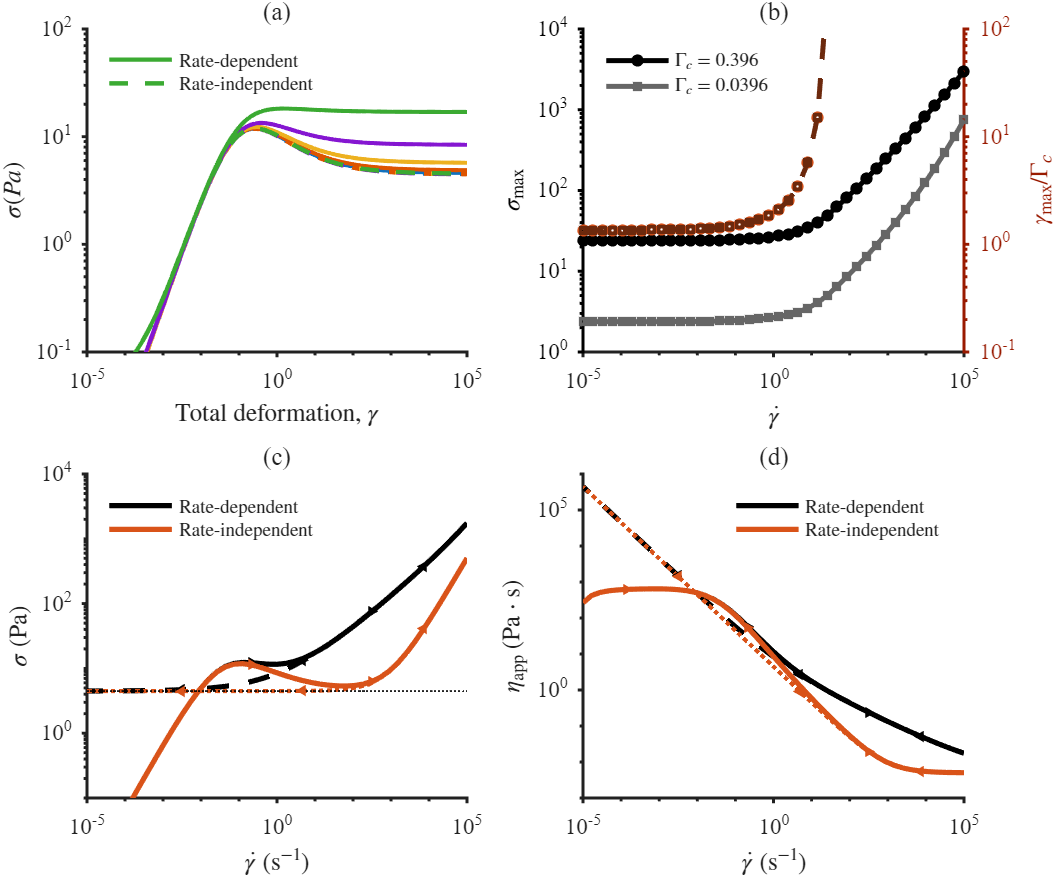}
\caption{\textbf{Transient rheological response of the load-transfer model.}(a) Startup stress overshoot as a function of irreversible deformation, $\Gamma$, for different shear rates (from $10^{-3} s^{-1}$ to $10^1 s^{-1}$). Solid and dashed curves correspond to rate-dependent $\Lambda_e=\Lambda_{e0}+K\dot{\gamma}^{1/2}$ and rate-independent interactions $\Lambda_e=\Lambda_{e0}$, respectively. (b) Dependence of the peak stress, $\sigma_{\max}$, and $\Gamma_{\max}/\Gamma_c$ on shear rate for  $\Gamma_c=0.396$, and $\Gamma_c=0.0396$. (c) Hysteresis during forward and backward shear-rate sweeps for rate-dependent and rate-independent interactions. (d) Apparent viscosity corresponding to (c) demonstrating the effect of timescale on it. Here, $\Gamma_L=0.01\Gamma_c$ is considered.}
\label{fig:fig3}
\end{figure}

\noindent Figure~\ref{fig:fig3}(a) demonstrates that the startup stress overshoot originates from the competition between elastic loading and deformation-induced structural degradation. Initially, elastic stress accumulates faster than the structure degrades, leading to stress growth. Beyond the overshoot, degradation progressively suppresses elastic stress storage and drives toward the frictionally saturated state. Rate-dependent interactions, $K\dot{\gamma}^n$, primarily modify the magnitude of the response while largely preserving the underlying overshoot mechanism. Figure~\ref{fig:fig3}(a) further shows that the overshoot weakens with increasing shear rate, and the response approaches a nearly monotonic saturation at high shear rates ($\dot{\gamma}\gtrsim10{\rm s^{-1}}$). The parameter $\Gamma_c$ represents the characteristic deformation scale associated with deformation-driven structural degradation. Thus, the location of the stress overshoot is largely controlled by $\Gamma_c$ [Fig.~\ref{fig:fig3}(b)], although the precise overshoot strain also depends on the interplay between elastic loading, structural rearrangement, and frictional stress. At low shear rates, $\Gamma_{max}\approx\Gamma_c+\Gamma_L$. As the shear rate increases, rapid loading enhanced rate-dependent frictional interaction, leading to an increase in both $\sigma_{\max}$ and $\Gamma_{max}/\Gamma_c$. These results suggest that $\Gamma_c$ controls the deformation scale associated with structural degradation and may be estimated experimentally from the overshoot location.

\noindent Figure~\ref{fig:fig3}(c)  shows upward and downward shear-rate sweeps with and without a rate-dependent interaction strength. While upward sweeps, $\dot{\gamma}=10^{-5} \rightarrow 10^{5}s^{-1}$ are performed in 20 steps with a $1{\rm s}$ waiting time at each shear rate, steady-state downward sweeps are considered. The upward sweep reflects the transient response of the material. For the rate-dependent model, the response evolves from a finite low-shear stress plateau through an extended shear-thinning regime to a viscous-dominated regime at high shear rates, consistent with the Herschel--Bulkley-like constitutive form derived in Eq.~(11). In contrast, the rate-independent interaction model reduces to $\sigma=\sigma_y+\eta_s\dot{\gamma}$,  corresponding to the classical Bingham model. The hysteresis between the upward and downward sweeps arises from the deformation induced softening and the strucrual load-transfer, which causes the transient startup response to differ from the steady-state constitutive behavior.

\noindent Figure~\ref{fig:fig3}(d) shows the apparent viscosity that is correlated with stress hysteresis shown in Fig.~\ref{fig:fig3}(c).  At short observation times, transient behavior, the response is characterized by an approximately
constant apparent viscosity, consistent with the  Barnes observation~\cite{Barnes1999}. As material evolves toward steady state, the apparent viscosity increases with decreasing shear rate, consistent with the observations of Møller~\cite{Moller2009}. The present framework provides a unified interpretation of these seemingly contrasting experimental observations by showing that they correspond to different stages of deformation-induced structural evolution. At sufficiently high shear rates, the apparent viscosity again approaches the constant solvent viscosity, recovering the Newtonian limit.  

\textit{Transient Shear Localization}: The emergence of a stress overshoot indicates a transient regime in which deformation-induced softening competes with elastic loading[Fig.~\ref{fig:fig3}(a)]. Such softening has been identified as an important mechanism underlying transient shear localization in complex fluids and soft glassy materials \cite{MoorcroftFielding2013,Fielding2014}. Recent linear stability analyses have demonstrated that transient flow instabilities can emerge from the evolving constitutive dynamics during startup, even in the absence of a stress overshoot \cite{Sharma2024}. In the present model, transient softening leads to $\frac{\partial \sigma}{\partial \Gamma}<0.$  The transient localization predicted by the present model appears to originate from deformation-induced elastic softening rather than from steady-state constitutive non-monotonicity or classical viscosity-bifurcation mechanisms associated with structural rebuilding \cite{Coussot2002}. A quantitative instability criterion would require a dedicated linear stability analysis and multidimensional simulations.

\textit{Discussion and Conclusions}: The central result of this work is that a finite yield stress emerges from structural load transfer. Furthermore, the long-standing debate regarding the existence of a yield stress together with the observation of time-dependent viscosity has been explained. In the proposed framework, increasing nonlinear deformation progressively degrades the load-bearing structure capable of storing recoverable nonlinear elastic energy while simultaneously promoting frictional interactions that approach a finite saturation limit. The resulting asymptotic frictional stress, together with the linear elastic stress, defines an emergent yield stress without invoking prescribed yield criteria, constitutive switching, or viscosity divergence. Yielding therefore appears not as a constitutive assumption but as a direct consequence of structural load transfer.

\noindent This perspective provides a unified interpretation of several rheological phenomena that are often treated separately. The yield stress originates from structural load transfer, whereas startup stress overshoots arise from the transient competition between elastic loading and deformation-induced structural degradation. Rate-dependent frictional interactions naturally generate Herschel--Bulkley shear thinning, while structural rebuilding introduces steady-state shear banding that are frequently described using distinct fluidity, aging, thixotropic, or thixo-elastoviscoplastic frameworks~\cite{Sollich1997,KumarL,Fluidity,Derec2001,Varnik2003,Coussot2002,Dimitriou2014}. 
Furthermore,the momentum-conserving analysis presented in the Supplementary Material shows that stress-controlled creep develops continuously from a quiescent state through material acceleration before the constitutive stresses progressively assume the stress balance. The model therefore suggests a different physical interpretation of yielding in disordered soft solids. Finally, because of emergent yielding, the framework naturally enables multidimensional simulations of plug formation and flow localization without constitutive regularization, constitutive switching, or explicit yield-surface tracking.

\section{ACKNOWLEDGMENTS}
L.K. acknowledges financial support from the Anusandhan National Research Foundation (ANRF) through the ARG-MATRICS program (ANRF/ARGM/2025/002005/MTR). L.K. also acknowledges support from the Science and Engineering Research Board (SERB), Government of India, through a Core Research Grant (CRG/2022/008946). L.K. is grateful to Prof. David A. Weitz for hosting and discussing the proposed model. L.K. also thanks Prof. R. Dasgupta and Prof. V. Mehandia for proofreading the manuscript and providing valuable comments and suggestions. The authors used generative artificial intelligence tools solely to assist with language editing and manuscript preparation. All scientific content, analysis, interpretations, and conclusions were developed, verified, and approved by the authors.

\end{document}